\documentclass[reprint, superscriptaddress, secnumarabic, amssymb, nobibnotes, aps, prl]{revtex4-1}

\usepackage{chemformula,siunitx}
\usepackage{lastpage}
\usepackage[protrusion=true, expansion=true]{microtype}
\usepackage{wrapfig}
\usepackage{multirow}
\usepackage{array}
\usepackage{ragged2e}
\justifying

\usepackage{graphicx}
\usepackage{epstopdf}
\usepackage[T1]{fontenc}
\usepackage{amsbsy}
\usepackage[T1]{fontenc}
\usepackage{amsmath}
\usepackage{amssymb}
\usepackage{bbm}
\usepackage{braket}
\usepackage{xcolor}
\allowdisplaybreaks
\usepackage{graphicx}

\usepackage[normalem]{ulem}

\newcommand{\equref}[1]{Eq.~(\ref{#1})}

\newcommand{\figref}[1]{Fig.~\ref{#1}}

\newcommand{\tableref}[1]{Table~\ref{#1}}

\usepackage[colorlinks=true]{hyperref}
\hypersetup{
    unicode=false,          
    pdftoolbar=true,        
    pdfmenubar=true,        
    pdffitwindow=false,     
    pdfstartview={FitH},    
    pdftitle={Ti1-xTaxSe2},    
    pdfauthor={Poulami Manna},      
    pdfsubject={},   
    pdfcreator={},   
    pdfproducer={}, 
    pdfkeywords={} {} {}, 
    pdfnewwindow=true,      
    colorlinks=true,       
    linkcolor=blue, 
    citecolor=blue,        
    filecolor=magenta,      
    urlcolor=blue          
}

\begin{document}

\title{\textrm{Quasi-two-dimensional superconductivity in 1$T$-Ti$_{1-x}$Ta$_x$Se$_2$}}

\author{P. Manna}
\affiliation{Department of Physics, Indian Institute of Science Education and Research Bhopal, Bhopal, 462066, India}
\author{S. Sharma}
\affiliation{Department of Physics, Indian Institute of Science Education and Research Bhopal, Bhopal, 462066, India}
\author{T. Agarwal}
\affiliation{Department of Physics, Indian Institute of Science Education and Research Bhopal, Bhopal, 462066, India}
\author{S. Srivastava}
\affiliation{Department of Physics, Indian Institute of Science Education and Research Bhopal, Bhopal, 462066, India}
\author{P. Mishra}
\affiliation{Department of Physics, Indian Institute of Science Education and Research Bhopal, Bhopal, 462066, India}
\author{R.~P.~Singh}
\email[]{rpsingh@iiserb.ac.in}
\affiliation{Department of Physics, Indian Institute of Science Education and Research Bhopal, Bhopal, 462066, India}

\begin{abstract}
The emergence of two-dimensional (2D) superconductivity in bulk transition metal dichalcogenides (TMDs) is a fascinating area of research, as their weak interlayer coupling leads to novel superconducting behavior and offers a rich platform to host nontrivial gap structures and interactions with other electronic orders. In this work, we present a comprehensive study of the superconducting properties of bulk single-crystalline $1T$-Ti$_{1-x}$Ta$_x$Se$_2$ for x = 0.2. Our results confirm the weakly coupled anisotropic superconductivity. Angle-dependent upper critical field measurements and observation of a Berezinskii-Kosterlitz-Thouless transition confirm the quasi-2D nature of the superconducting state. These results position $1T$-Ti$_{1-x}$Ta$_x$Se$_2$ as a promising platform for exploring low-dimensional superconducting physics and highlight bulk TMD crystals as a promising platform for realizing intrinsic 2D superconductivity, opening avenues for future quantum applications.

\end{abstract}
\maketitle
\section{INTRODUCTION}
Superconductivity in two-dimensional (2D) systems has garnered widespread attention for hosting exotic quantum states \cite{lian2023interplay, qiu2021recent, tang2025preparation, gantmakher2010superconductor, PhysRevB.100.064506, si2021elemental, xi2016ising, choe2016understanding, m9xx-gk46} and has become the perfect avenue for cutting-edge device applications \cite{Kaul_2014}. Layered transition-metal dichalcogenides (TMDs), superconducting thin films, ion-gated superconductors, and exfoliated 2D crystals are key platforms for these kinds of investigations \cite{PhysRevLett.86.4382,baturina2012superconducting, saito2016gate, saito2016highly}. 2D superconductors remain a focus of intense research due to their remarkable characteristics, particularly the violation of the Pauli limit and the Berezinskii-Kosterlitz-Thouless (BKT) transition. Interestingly, in the atomic layer limit, superconducting 2$H$-TaS$_2$ and 2$H$-NbSe$_2$ showcase an enhanced in-plane upper critical field, surpassing the Pauli limit, attributed to the Ising pairing driven by valley-dependent spin-orbit coupling (SOC) and broken inversion symmetry \cite{de2018tuning, li2021printable}. The BKT mechanism, another hallmark of 2D nature, occurs when vortex-antivortex pairs form, resulting in power-law behavior in the current-voltage characteristics. A recent report on clean 2D superconductivity in a bulk superlattice of Ba$_6$Nb$_{11}$S$_{28}$ presents an example of reduced dimensionality \cite{devarakonda2020clean}, indicating that layered superconductors with a variety of 2D characteristics can be generated by sufficiently weakening the interlayer coupling. Given the experimental difficulties in fabricating monolayer systems, exploring 2D superconductivity in bulk crystals offers a more accessible and practical alternative. Introducing intercalation, chemical doping, or insulating layers into layered materials or making heterostructures presents a viable approach to achieve 2D superconductivity by significantly weakening the interlayer coupling, \cite{PhysRevMaterials.4.124803, lu2015evidence, li2019phase, patra2022two, agarwal2023quasi, shi2024two, fan2025two, gmcz-mw9g}. Chemical doping has an advantage over intercalation, given its significant control over carrier density and structural stability, and is suitable for easy exfoliation and device fabrication \cite{kim2016metallic, Hu_2021}. 

While 2D superconductivity has been observed in doped bulk crystals of 2$H$-TaS$_2$ and 2$H$-NbSe$_2$ \cite{patra2022two, patra2024planar}, Ti-based dichalcogenides remain largely unexplored. Among them, $\text{TiSe}_2$ is extensively studied due to its controversial electronic nature and semimetallic/ semi-semiconducting behavior \cite{friend1977semimetallic, rasch2007electronic, rasch20081}. At ambient pressure, $\text{TiSe}_2$ hosts a commensurate CDW phase at 200 K \cite{wegner2018evidence, di1976electronic}, but does not exhibit superconductivity \cite{morosan2010multiple}. However, under pressure of around 3 GPa, it exhibits a superconducting transition at 1.8 K, accompanied by CDW suppression \cite{kusmartseva2009pressure}. Interest in $\text{TiSe}_2$ grew after the discovery of superconductivity concurrent with the meltdown of the CDW in Cu-intercalated 1$T$-TiSe$_2$. Muon-spin rotation on Cu$_{0.08}$TiSe$_2$ suggested a two-gap superconducting state \cite{zaberchik2010possible}, while the surface superconductivity in Cu$_{0.1}$TiSe$_2$ \cite{levy2016puzzling} prompted further exploration with other intercalants. In addition to Cu, intercalation by Pd, H, and PbSe layers in TiSe$_2$ also induces superconductivity \cite{morosan2010multiple, piatti2025direct, PhysRevB.82.024503}. Recent reports demonstrate that Ta doping enhances the density of states near the Fermi level, leading to suppression of the CDW phase and the emergence of superconductivity \cite{Hu_2021, PhysRevB.110.165156, luo2016differences}. Similarly to other doped TMDs, doping can also induce 2D superconductivity in $\text{TiSe}_2$, making it a promising candidate to understand the superconducting nature and pairing symmetry in low-dimensional limits.

In this work, we report the synthesis of the crystalline 1$T$-Ti$_{1-x}$Ta$_x$Se$_2$ (for x = 0, 0.05, 0.1, 0.2, 0.3). The composition x = 0.2 exhibits the maximum transition temperature of $T_c$ = 2.32 (1) K. A detailed investigation of its superconducting properties is conducted via magnetization, specific heat, and AC transport measurements, suggesting weakly coupled type-II anisotropic superconductivity. Furthermore, angle-dependent magneto-transport measurements and observation of a BKT transition provide the first evidence of quasi-2D superconductivity in a Ti-based TMD system.

\begin{figure*}
\includegraphics[width=1.99\columnwidth]{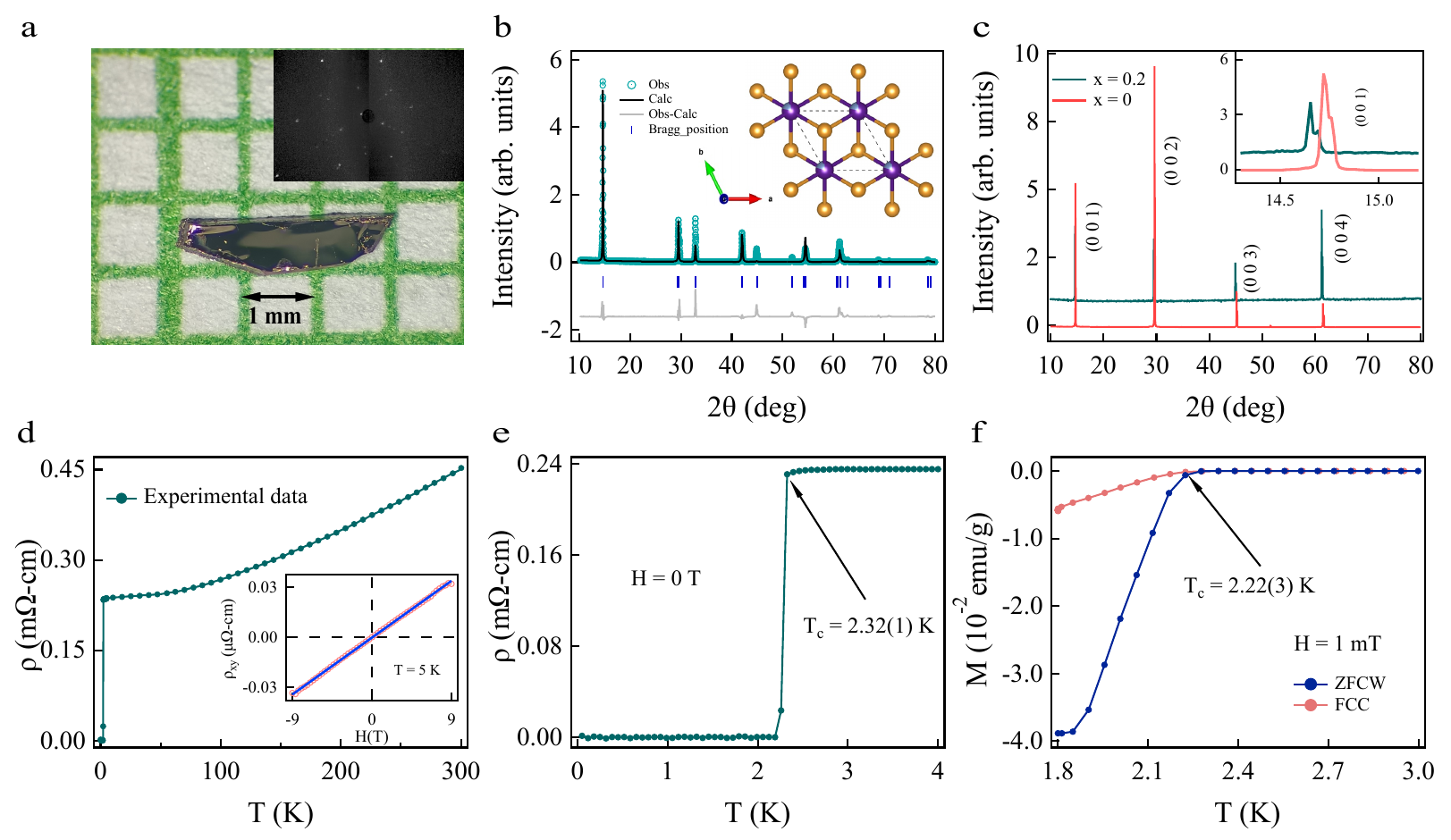}
\caption {\label{fig1}(a) A single crystal-image with Laue spots in the inset. (b) Rietveld refined powder XRD patterns. The Bragg position, theoretical refinement, and experimental results are symbolized in that sequence by marks, lines, and vertical bars, respectively. The difference between the computed and experimental data is indicated by the line at the bottom. Inset: Crystal structure of 1$T$-Ti$_{0.8}$Ta$_{0.2}$Se$_2$ is viewed along the out-of-plane direction. Purple, brown, and blue spheres correspond to Ti, Se, and Ta atoms, respectively.  (c) Single-crystal XRD patterns of undoped and Ta-doped TiSe$_2$, represented by pink and teal colors, respectively. Inset: The observed shift corresponds to a change in the 2$\theta$. (d) Temperature-dependent resistivity at zero field. Inset shows the Hall resistivity with positive slope at T = 5 K. (e) Abrupt drop in resistivity occurs at $T$$_{c,}$$_{onset}$ = 2.32(1) K. (f) Magnetization versus temperature measurements show superconducting transition at a temperature of 2.22(3) K.}
\end{figure*}

\section{EXPERIMENTAL DETAILS}
Single-crystals of Ti$_{1-x}$Ta$_x$Se$_2$ were synthesized utilizing the standard chemical vapor transport (CVT) method using iodine (I$_2$) as a transport agent. High-purity Ta (99.97\%), Ti (99.99\%), and Se (99.999\%) powders in precise stoichiometric ratios were ground and sealed in an evacuated quartz ampule together with I$_2$ (5 mg/cc). The ampule was then placed in a tubular furnace, where a temperature gradient of 50 K was applied. The hot zone was maintained at a temperature of 998 K. After 10 days, the ampoule was quenched in ice water, resulting in the formation of large shiny copper-colored crystals in the cold zone. The structure and phase purity of the compound were analyzed using powder X-ray diffraction (PXRD) conducted on an X'pert PANalytical Empyrean X-ray diffractometer with monochromatic Cu-$K_\alpha$ radiation ($\lambda$ = 1.54 \AA). Rietveld refinement was performed on the powder diffraction data using the FULLPROF suite software. Elemental compositions were verified through energy-dispersive X-ray analysis (EDAX) performed with scanning electron microscopy (SEM). Magnetization measurements were performed using the Quantum Design magnetic property measurement system (MPMS3), equipped with a $^4He$ cryostat. Transport measurements and specific heat measurements were performed using a 9T Quantum Design physical property measurement system (PPMS) and a Dilution Refrigerator (DR).

\begin{figure*}
\includegraphics[width=2.00\columnwidth]{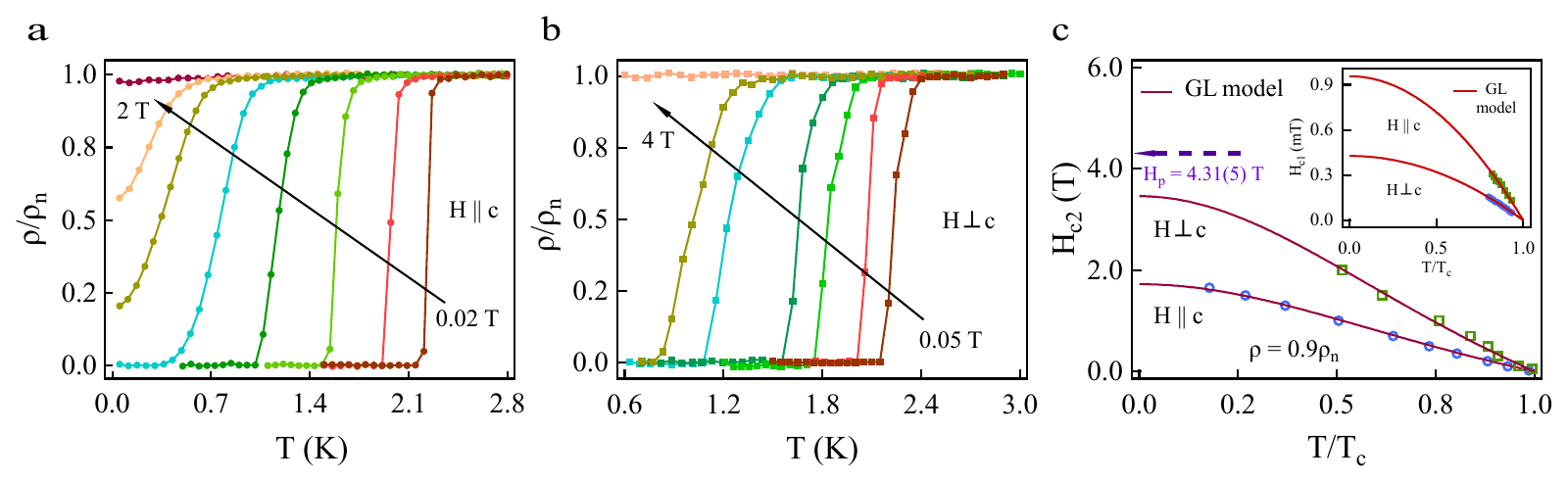}
\caption {\label{fig2} (a) and (b) Temperature-dependent normalized electrical resistivity for $H \parallel c$ and $H\perp c$ with different magnetic fields, respectively. (c) Upper critical fields in both directions are also fitted by the GL model as a function of temperature. The Pauli limit is 4.31(5) T, as shown by the purple dotted lines. Inset: Temperature-dependent lower critical fields in both directions. GL-fitting is shown by the red lines.}
\end{figure*}

\section{RESULTS AND DISCUSSION}
\subsection{Sample characterization}
\figref{fig1}(a) shows the microscopic image of a shiny and as-grown single crystal.  The Laue diffraction pattern (inset of \figref{fig1}(a)) validates the single-crystalline nature of the sample. The powder XRD patterns of the crushed crystal of Ti$_{0.8}$Ta$_{0.2}$Se$_2$ were analyzed by Rietveld refinement \cite{fullprof}, which is shown in \figref{fig1}(b). The corresponding refinement confirms the single phase of the crystal. Similar to the pristine compound, this doped compound has a trigonal structure (1$T$) with the $ P\bar3m1$ space group (164). The refined lattice parameters are $a$ = $b$ = 3.5252(3) $\text{\AA}$, $c$ = 6.0490(8) $\text{\AA}$, where the parameter $c$ shows a considerable increase, indicating weak interlayer coupling. The lattice parameters for TiSe$_2$ and Ta-doped TiSe$_2$ are compared in \tableref{tbl1}. The inset of \figref{fig1}(b) presents the crystal structure of Ti$_{0.8}$Ta$_{0.2}$Se$_2$, generated by VESTA software \cite{momma2011vesta}. \figref{fig1}(c) show single crystal XRD patterns for TiSe$_2$ and Ta-doped TiSe$_2$, where the crystals are oriented along the $(00n)$ direction, indicating excellent $c$-axis characteristics. Comparison of single-crystal XRD patterns implies a shift to a lower angle by Ta doping, clearly reflected in the inset of \figref{fig1}(c). Fig. \textcolor{blue}{S1}(a) and (b) of the Supplemental Material present the shift in 2$\theta$ for various Ta-dopants \cite{Supp}. Furthermore, EDAX measurements performed in various regions of the single crystal revealed a consistent average elemental composition (see Supplemental Material \cite{Supp}, Fig. \textcolor{blue}{S1}(c)).

\begin{table}[b]
\caption{Lattice parameters and transition temperature of Ti$_{0.8}$Ta$_{0.2}$Se$_2$ compared with the parent compound.}
\label{tbl1}
\setlength{\tabcolsep}{7pt}
\begin{center}
\begin{tabular}[b]{lccc}\hline
&Structure    &Trigonal  \\
&Space group  &$P\bar3m1$  (164) \\
      \hline
      & $a$ = $b$ (\text{\AA}) & $c$ (\text{\AA}) & $T_c$ (K) \\
\hline                             
TiSe$_2$ \cite{kitou2019effect} & 3.5395(3) & 6.0082(4) & - \\
Ti$_{0.8}$Ta$_{0.2}$Se$_2$ & 3.5252(3) & 6.0490(8) & 2.32(3) \\
\hline
\end{tabular}
\par\medskip\footnotesize
\end{center}
\end{table}

\subsection{Anisotropic superconductivity}
The temperature-dependent electrical resistivity of crystalline Ti$_{0.8}$Ta$_{0.2}$Se$_2$  exhibits a sharp drop to zero at a specific temperature when measured in a zero magnetic field over the range of 0.05 K to 300 K (depicted in \figref{fig1}(d)). The gradual decrease in resistivity from 300 K to 5 K signifies the metallic nature of the material. The residual resistivity ratio (RRR) (denoted as $\rho(300 K)$/$\rho(5 K)$) is approximately 2. The onset superconducting transition temperature $T$$_{c,}$$_{onset}$ = 2.32(1) K, which is shown in \figref{fig1}(e). Unlike in the parent compound, no anomaly was detected in the normal-state resistivity around 200 K, pointing out the absence of the CDW phase, likely due to the incorporation of the 5d element Ta. This behavior is well supported by previous reports. A comparison of temperature-dependent electrical resistivity for single crystals of Ti$_{1-x}$Ta$_x$Se$_2$ is shown in Supplemental Material \cite{Supp}, Fig. \textcolor{blue}{S3}(a).

Hall resistivity measurement has been used to determine the carrier density of Ti$_{0.8}$Ta$_{0.2}$Se$_2$. As can be seen in the inset of \figref{fig1}(d), the slope value generates the Hall coefficient ($R_H$), where the Hall resistivity $\rho_{xy}$ exhibits a general linear behavior with changing field. $R_H$ = 10.6 $\times$ 10$^{-12}$ cm$^3$C$^{-1}$ was obtained by linear fit. The holes are the predominant charge carriers of the system, as reflected in the positive value of $R_H$. The carrier density $n$ = 5.91(4) $\times$ 10$^{27}$ m$^{-3}$ was derived using the formula: $R_H$ = 1/$n$e. Using $n$, we also calculated the electronic parameters, which are listed in \tableref{tbl3} (detailed calculations are provided in the Supplemental Material \cite{Supp}).

\figref{fig1}(f) displays the temperature-dependent magnetization obtained from two distinct modes: zero-field-cooled warming (ZFCW) and field-cooled cooling (FCC) under an applied magnetic field of 1 mT. The diamagnetic signal observed in these modes suggests the presence of bulk superconductivity at 2.22(3) K.

Field-dependent magnetization curves were analyzed to calculate the lower critical field ($H_{c1}$(0)) by extracting the $H_{c1}$ values at a fixed temperature, identified at the point where the M-H curves deviate from the linear Meissner response, as illustrated in Supplemental Material \cite{Supp}, Fig. \textcolor{blue}{S2}. $H_{c1}$(0) for both directions was determined by fitting the retrieved values $H_{c1}$ as a function of reduced temperature ($\frac{T}{T_{c}}$) using the conventional Ginzburg-Landau (GL) relation defined as
\begin{equation}
H_{c1}(T)=H_{c1}(0)\left[{1-t^{2}}\right], \quad  \text{where} \;  t = \frac{T}{T_{c}}.
\label{eqn3:HC1}
\end{equation}
which gives $H_{c1}$(0) = 0.96(1) and 0.43(3) mT for $H$ parallel to the $c$-axis and $H$ perpendicular to the $c$-axis, respectively, as given in the inset of \figref{fig2}(c).

To estimate the upper critical field values ($H_{c2}$(0)) for both field orientations, temperature-dependent resistivity measurements were carried out under various applied magnetic fields (\figref{fig2}(a) and (b) for $H\parallel c$ and $H\perp c$, respectively). As the magnetic field increased, the superconducting transition temperature ($T_{c}$) systematically decreased. The values of $H_{c2}$ were extracted from the resistivity curves by defining the criterion $\rho$ = 0.9$\rho_n$, where $\rho_n$ denotes the resistivity of the normal-state. The extrapolated $H_{c2}$ values were then fitted as a function of reduced temperature ($\frac{T}{T_{c}}$) using the GL-equation:
\begin{equation}
H_{c2}(T) = H_{c2}(0)\left[\frac{1-t^{2}}{1+t^{2}}\right],  \quad  \text{where} \;  t = \frac{T}{T_{c}}.
\label{eqn4:HC2}
\end{equation}
The $H_{c2}$ (0) values obtained from the GL adjustment are 1.72(6) and 3.46(3) T for $H \parallel c$ and $H \perp c$, respectively, as given in \figref{fig2}(c). This clear directional dependence of $H_{c2}$(0) highlights the anisotropic nature of the compound, with an anisotropy ratio of about 2. A comparable degree of anisotropy ($\thicksim$ 1.7) has also been achieved in Cu$_{0.1}$TiSe$_2$ \cite{kavcmarvcik2013heat}, which can be attributed to the anisotropy of its Fermi surface and quasi-two-dimensional crystal structure.

The suppression of superconductivity by a magnetic field is primarily caused by two key mechanisms: the Pauli limiting effect and the orbital limiting effect. The former breaks Cooper pairs via Zeeman splitting by aligning the spins of electrons in the same direction, whereas the orbital-limiting effect disrupts the continuous flow of Cooper pairs, leading to the formation of vortices. These combined effects define the upper critical field beyond which superconductivity can no longer exist. According to BCS theory, the Pauli limit is given by $H_{c2}^{P}$(0) = 1.86 $T_{c}$ \cite{Chandrasekhar1962pauli, Clogston1962pauli}. $H_{c2}^{P}$(0) is calculated as 4.31(5) T for Ti$_{0.8}$Ta$_{0.2}$Se$_2$ superconductor, taking $T_{c}$ = 2.32(1) K. This value is above the in-plane upper critical field, implying no violation of the Pauli limit. The orbital-limiting effect is evaluated using the Werthamer-Helfand-Hohenberg (WHH) theory for type-II superconductors, assuming negligible spin-orbit coupling \cite{WHH1966orbital, Helfand1966orbital}. It is defined as 
\begin{equation}
H^{orb}_{c2}(0) = -\alpha T_{c} \left.{\frac{dH_{c2}(T)}{dT}}\right|_{T=T_{c}}, 
\label{eqn5:WHH}
\end{equation}
where $\alpha$ is a constant, known as the purity factor, taking values of 0.69 and 0.73 for superconductors in the dirty and clean limit, respectively. The estimated value of $H^{orb}_{c2}(0)$ for $H\perp c$ is 1.99(2) T for $\alpha$ = 0.69. As the upper critical field values for both orientations are significantly lower than the Pauli limiting field, the orbital effect is likely responsible for Cooper pair breaking.
\begin{table}[b]
\caption{The anisotropic superconducting parameters of the synthesized single crystal.}
\label{tbl2}
\setlength{\tabcolsep}{12pt}
\begin{center}
\begin{tabular}[b]{lccc}\hline
Parameters& Unit & $H\parallel c$ & $H\perp c$ \\
\hline
$H_{c1}(0)$ & mT & 0.96(1) &   0.43(3) \\ 
$H_{c2}^{res}$(0) & T & 1.72(6) &   3.46(3) \\
$\xi$ & nm & 6.90(1) &  13.8(4)  \\
$\lambda_{GL}$ & nm & 838.21(8)  &2265.4(2)  \\
$\kappa_{GL}$& & 74.1 & 141.1  \\
\hline
\end{tabular}
\end{center}
\end{table}
The correlation between the coherence length ($\xi$) and the upper critical field for anisotropic superconductors can be stated with \equref{eqn4: coherence} as follows \cite{palstra1988angular}:
\begin{equation}
    H_{c2} = \frac{\phi_0}{2\pi\xi_{\perp c}^2}(cos^2\theta+\epsilon^2sin^2\theta)^{-1/2}
    \label{eqn4: coherence}
\end{equation} 
where $\phi_{0}$ = h/2e the magnetic flux quanta have a value of 2.07$\times$10$^{-15}$ T$m^{2}$. $\epsilon$ is the ratio of the two coherence lengths, while $\theta$ is the angle between the applied field and the unit vector normal to the layers,  i.e., $\epsilon$ = $\xi_{\parallel c}$/$\xi_{\perp c}$. We can assess the formulas for the coherence length along the parallel ($\xi_{\parallel c}$) and perpendicular direction ($\xi_{\perp c}$) to the x-axis by reducing \equref{eqn4: coherence} for $\theta = 0^{\circ}$ and 90$^{\circ}$. So, the formulated expressions are $H_{c2}^\parallel(0)$ = $\frac{\phi_0}{2\pi\xi^2_{\perp c}}$ and $H_{c2}^\perp(0)$ = $\frac{\phi_0}{2\pi\xi_{\parallel c}\xi_{\perp c}}$. Using the above equations, the values obtained for $\xi_{\parallel c}$ and $\xi_{\perp c}$ are 6.9(1) nm and 13.8(4) nm, respectively.

GL penetration length and GL parameter $\kappa$ were acquired using a set of standard equations: $H_{c2}^\perp$(0)/$H_{c1}^\perp$(0) = 2$\kappa_{\perp c}^2/ln \kappa_{\perp c}$, $\kappa_{\perp c}$ = [$\lambda_{\perp c}$(0)$\lambda_{\parallel c}$(0)/$\xi_{\perp c}$(0)$\xi_{\parallel c}$(0)]$^{1/2}$ and $\kappa_{\parallel c}$ = $\lambda_{\perp c}$(0)/$\xi_{\perp c}$(0). \tableref{tbl2} gives an overview of all the measured anisotropic superconducting parameters. Since $\kappa$> 1/$\sqrt{2}$, Ti$_{0.8}$Ta$_{0.2}$Se$_2$ has appeared as a strong type-II superconductor. The thermodynamic field, which reflects the superconducting condensation energy, was calculated to be roughly 0.02 T with $H_c(0) = H_{c1}^\perp(0)\sqrt{2}\kappa_{\perp c}/ln \kappa_{\perp c}$.

\begin{figure*}
\includegraphics[width=1.5\columnwidth]{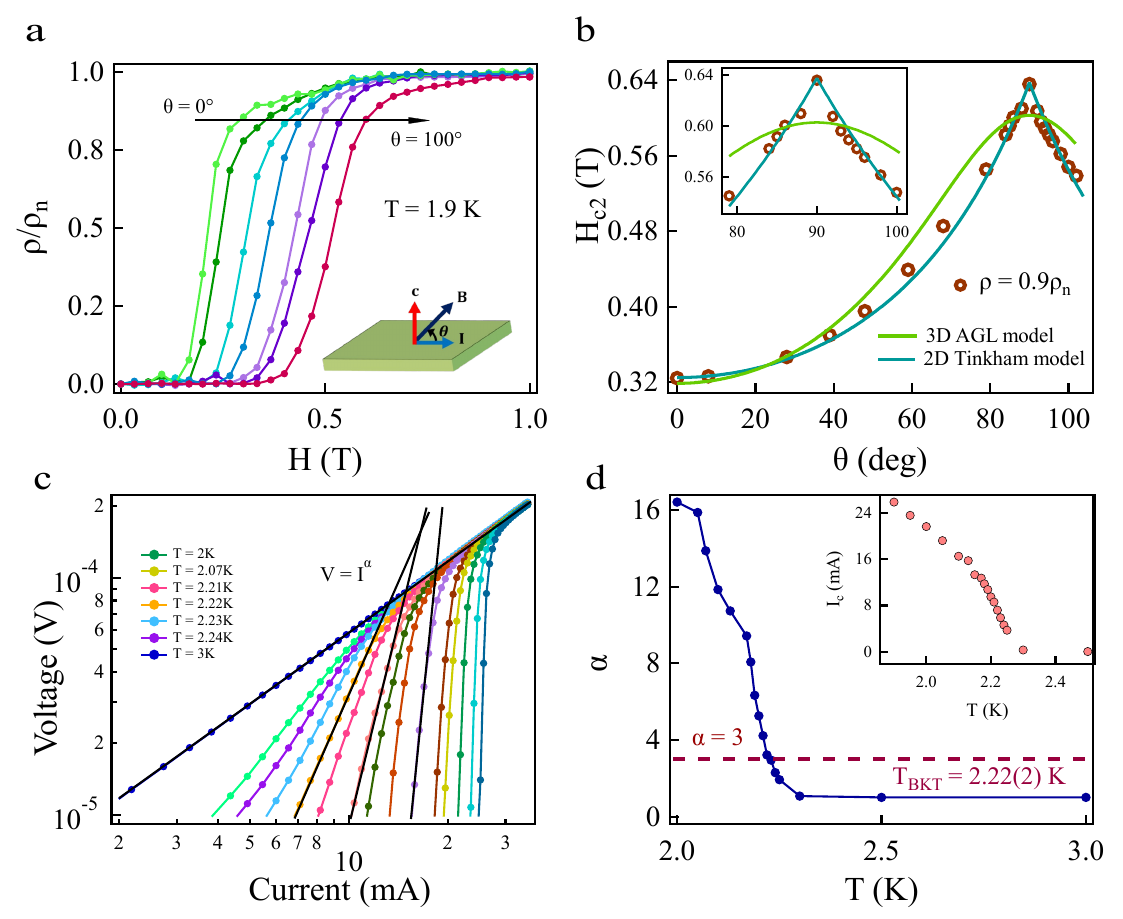}
\caption {\label{fig3}(a) Field-dependent electrical resistivity with various angles shows anisotropy in this sample. (b) Upper critical fields versus angle for $\rho = 0.9\rho_n$ are best described by the two-dimensional Tinkham model rather than the three-dimensional anisotropic GL model. Inset: Expanded view of the fitting near 90$^{\circ}$ angle. (c) The temperature-dependent $V$-$I$ curves are shown on a logarithmic scale. (d) The evolution of the slopes of the $V$-$I$ curves, identifying the BKT jump in $\alpha (T)$ at 2.22(2) K. The slope of $\alpha (T)$ = 3 is marked by the purple dotted line. Inset: The variation of critical current $I_c$ with temperature.}
\end{figure*}

\subsection{Two-dimensional superconductivity}

\textbf{Angle-dependent upper critical field.} 
Identification of anisotropy in superconductors, along with 2D superconducting properties, is aided by the angle-dependent variation of the upper critical field. At a fixed temperature of 1.9 K, the field-dependent resistivity was measured at several angles, as illustrated in the \figref{fig3}(a), where $\theta$ is the angle between the magnetic field and the normal to the sample plane. Setting $\rho$ = 0.9$\rho_n$, the upper critical fields are extracted for individual angles. As seen in \figref{fig3}(b), a pronounced cusp is observed near $\theta$ = 90$^{\circ}$. The enlarged view of $H_{c2}$ versus $\theta$ is provided in the inset of \figref{fig3}(b). Generally, there are two theoretical models that explain the angle-dependent behavior of the upper critical field. In the case of three-dimensional (3D) superconductors, the anisotropic GL (AGL) model (\equref{eqnAGL}) gives the ellipsoidal form of $H_{c2}$, whereas Tinkham proposed a model for 2D thin-film superconductors (\equref{eqn2DT}).
\begin{equation}\label{eqnAGL}
    \left(\frac{H_{c2}(\theta,T) \sin{\theta}}{H_{c2}^{\perp}}\right)^2 + \left(\frac{H_{c2}( \theta,T) \cos{\theta}}{H_{c2}^{||}}\right)^2 = 1
\end{equation}

\begin{equation}\label{eqn2DT}
    \left(\frac{H_{c2}(\theta,T) \sin{\theta}}{H_{c2}^{\perp}}\right)^2 + \left\arrowvert\frac{H_{c2}( \theta,T) \cos{\theta}}{H_{c2}^{||}}\right\arrowvert = 1
\end{equation}
The data fitted using the 2D Tinkham model gives a better result compared to the 3D AGL model \cite{patra2024planar}. Furthermore, for the resistivity criteria of $\rho$ = 0.5$\rho_n$ and $\rho$ = 0.1$\rho_n$, the data remain consistent with the 2D model as shown in Supplemental Material \cite{Supp}, Fig. \textcolor{blue}{S3}(b). These observations point to a possible quasi-2D nature of the superconductor 1$T$-Ti$_{0.8}$Ta$_{0.2}$Se$_2$. 

\quad

\textbf{Berezinskii-Kosterlitz-Thouless (BKT) transition.}
The Berezinskii-Kosterlitz-Thouless (BKT) transition typically emerges as the system evolves from a quasi-long-range ordered vortex state to a fully disordered phase, with increasing temperature \cite{takiguchi2024berezinskii}. This reflects the intricate interplay between order and disorder within the vortex condensate. The quasi-2D nature of Ti$_{0.8}$Ta$_{0.2}$Se$_2$ is further reinforced by the temperature-dependent current-voltage ($I$-$V$) characteristics (as shown in \figref{fig3}(c)), which reveal the BKT transition temperature. Below $T_c$, high currents can unbind vortex-antivortex pairs, leading to a non-ohmic response. In the BKT scenario, the voltage follows a power-law dependence (\equref{eqnV-I}), with the exponent proportional to $J_s$, the superfluid density. This behavior comes from the fact that the equilibrium density of the free vortices $n_v$($I$) scales with a power law of the applied current, thus contributing to the voltage response according to $V \propto$ $n_v$($I$)$I$ \cite{PhysRevB.100.064506}.
\begin{equation}
    V \propto I^{\alpha(T)},  
    \quad \alpha(T) = 1 + \pi \frac{J_s(T)}{T} .
    \label{eqnV-I}
\end{equation}
In the ideal case of a BKT transition, it is expected that the superfluid density $J_s$ undergoes a discontinuous jump at the point where it intersects the BKT line.
\begin{equation}
    J_s(T^-_{BKT}) = \frac{2}{\pi T_{BKT}},  
    \quad J_s(T^+_{BKT}) = 0.
    \label{eqnBKT}
\end{equation}
Substituting this into \equref{eqnV-I} implies that the V-I exponent should exhibit a jump at the transition:
\begin{equation}
    \alpha(T^-_{BKT}) = 3,  
    \quad\alpha(T^+_{BKT}) = 1.
    \label{eqnalpha}
\end{equation}
By fitting the power-law behavior into equation \equref{eqnV-I}, we extracted the exponent values as a function of temperature (depicted in \figref{fig3}(d)).  The resulting BKT transition temperature is estimated to be approximately 2.22(2) K. The inset of \figref{fig3}(d) shows the nearly steplike temperature-dependence of the critical current ($I_c$), obtained by extrapolating the linear region of the $V$-$I$ curves \cite{10.1063/1.4901940}. Observation of a BKT transition provides compelling evidence for the quasi-2D behavior of the superconductor studied.

\subsection{Specific heat and electronic parameters}
\begin{figure}
\includegraphics[width=0.96\columnwidth]{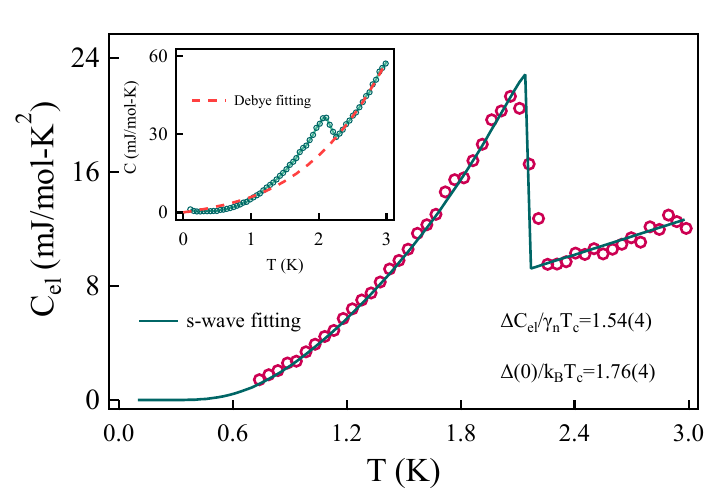}
\caption{\label{fig4} The electronic specific heat data are fitted by an isotropic s-wave model, rendering conventional superconductivity. Inset: Normal-state specific heat data are fitted using the Debye-Sommerfeld model.}
\end{figure}

The superconducting transition is further corroborated by a distinct anomaly in temperature-dependent specific heat data in the absence of a magnetic field. This results in bulk superconductivity at a temperature of 2.15 K in layered 1$T$-Ti$_{0.8}$Ta$_{0.2}$Se$_2$, which is in better agreement with the transition observed in both magnetization and resistivity measurements. Normal-state specific data were fitted using the Debye-Sommerfeld model (depicted in the inset of \figref{fig4}) described as $C$ = $\gamma_{n}T$ + $\beta_{3}T^{3}$, where $\gamma_{n}T$ tells about the electronic contribution to specific heat and $\beta_{3}T^{3}$ accounts for the phononic contribution at low temperatures. Here, $\gamma_{n}$ is the Sommerfeld coefficient and $\beta_{3}$ is the Debye constant. Using the expression above to fit the specific heat data from the normal state, the evaluated parameters are $\gamma_{n}$ = 4.26(1) mJmol$^-1$K$^{-2}$ and $\beta_{3}$ = 1.69(4) mJmol$^-1$K$^{-4}$. The Debye constant is used to calculate the Debye temperature $\theta_{D}$, which characterizes the phonon spectrum and vibrational properties of a crystalline solid. It is determined using the following relation: 
\begin{equation}
\theta_{D} = \left(\frac{12\pi^{4} R N}{5 \beta_{3}}\right)^{\frac{1}{3}},
\label{eqn8:DebyeTemperature}
\end{equation}
where, $R$ = 8.314 J mol$^{-1}$ K$^{-1}$ is a gas constant, 
$N$ = 3 number of atoms per formula unit. The value of $\theta_{D}$ is about 150.98(2) K. 

In BCS-type superconductors, the density of states at the Fermi level plays a vital role in understanding the pairing mechanisms. It can be estimated by incorporating $\gamma_{n}$, which is directly proportional to $N(E_{F})$. The formula $\gamma_{n}$ = $\left(\frac{\pi^{2} k_{B}^{2}}{3}\right) N(E_{F})$, where $k_{B}$ = 1.38 $\times$ 10$^{-23}$ J K$^{-1}$, is used to evaluate $N(E_{F})$. The resulting value of $N(E_{F})$ is 1.80(6) states eV$^{-1}$f.u.$^{-1}$. The electron-phonon coupling, denoted by $\lambda_{e-ph}$, quantifies the strength of the interaction between electrons and phonons. It can be modulated using modified McMillan's equation \cite{mcmillan1968transition}, stated as:
\begin{equation}
\lambda_{e-ph} = \frac{1.04+\mu^{*}\mathrm{ln}(\theta_{D}/1.45T_{c})}{(1-0.62\mu^{*})\mathrm{ln}(\theta_{D}/1.45T_{c})-1.04 };
\label{eqn9:Lambda}
\end{equation}
where $\mu^{*}$, a Coulomb pseudopotential, is usually taken as 0.13 for transition metals. Using this approach, the resulting $\lambda_{e-ph}$ is 0.61(1), for $T_{c}$ = 2.15 K and $\theta_{D}$ = 150.98(2) K, implying that the material is a weakly coupled BCS superconductor. 

\begin{table}[b]
\caption{The estimated normal state parameters.}
\label{tbl3}
\setlength{\tabcolsep}{10pt}
\begin{center}
\begin{tabular}[b]{lcc}\hline
Parameters& Unit \\
\hline
$\gamma_{n}$& mJ/mol K$^{2}$&  4.26(1) \\
$\theta_{D}$& K&150.98(2) \\
$N(E_{F})$ & states/(eV f.u.) & 1.80(6) \\ 
$m^{*}/m_{e}$& - &3.24(5) \\
$n$&10$^{27}$ $m^{-3}$&5.91(4) \\
$v_{F}$&10$^{5}$ m/s&1.99(4) \\
$\xi_{0}$&10$^{-7}$ m &1.18(2) \\
$l_{e}$&10$^{-9}$ m&1.68(7) \\
\hline
\end{tabular}
\end{center}
\end{table}

The relation between electronic-specific heat (C$_{el}$) and temperature provides valuable information on the superconducting pairing mechanism. The electronic contribution to the specific heat is obtained by subtracting the phononic contribution from the total zero-field-specific heat using the formula $C_{el} = C - \beta_{3}T^{3}$. Temperature-dependent $C_{el}$ is then plotted and analyzed using a single-gap s-wave model \cite{padamsee1973quasiparticle} as shown in \figref{fig4}. Within this framework, the entropy $S$ can be evaluated as
\begin{equation}
\begin{split}
\frac{S}{\gamma_{n} T_{c}}= -\frac{6}{\pi^{2}} \left(\frac{\Delta(0)}{k_{B} T_{C}}\right) &\int_{0}^{\infty}[ fln(f)\\
&+(1-f)ln(1-f)] dy,
\end{split}
\label{Eq:swave}
\end{equation}
where, $f(\xi)$ = $[e^{\beta E(\xi)}+1]^{-1}$ is the Fermi function, $E(\xi) = \sqrt{\xi^{2}+\Delta^{2}(t)}$, $\xi$ is the normal-electron energy and $\Delta(t)$ is the temperature-dependent gap function. The integration variable $y$ is defined as $\xi/\Delta(0)$. Using the isotropic s-wave BCS approximation, the superconducting gap function can be derived as $\Delta(t) = tanh[1.82(1.018[(1/t)-1])^{0.51}]$. The expression $C_{el}$ = $tdS/dt$ shows the correlation between the electronic specific heat ($C_{el}$) and the entropy ($S$). The gap ratio $\Delta(0)/k_{B}T_{C}$  of 1.76(4), which aligns well with the expected value for a weakly coupled BCS superconductor, is computed by fitting the temperature-dependent electron-specific heat data using the typical s-wave model. Additionally, the specific heat jump $\Delta C_{el}/\gamma_n T_c$ has been shown to be almost 1.54(4), comparable to the BCS value.

\section{Conclusion}

Single crystals of 1$T$-Ti$_{0.8}$Ta$_{0.2}$Se$_2$ were successfully synthesized by the chemical vapor transport method. Detailed magnetization, resistivity, and specific heat analyzes establish it as a bulk type-II anisotropic superconductor with a transition temperature $T_c$ = 2.32(1) K, accompanied by suppression of the charge density wave (CDW) transition. Specific heat results reveal weakly coupled superconductivity characterized by an isotropic s-wave energy gap, consistent with conventional BCS theory. Crucially, the quasi-two-dimensional nature of superconductivity is confirmed by angle-dependent upper critical field measurements fitting the 2D Tinkham model and the detection of a Berezinskii-Kosterlitz-Thouless (BKT) transition. This represents the first demonstration of quasi-2D superconductivity in Ti-based transition metal dichalcogenides. These findings open avenues for investigating 1$T$-TiSe$_2$ in the few-layer limit and exploring low-dimensional quantum phases prevalent in bulk TMDs with weak interlayer coupling, significantly advancing the pursuit of two-dimensional superconductivity.
\section{Acknowledgments} 
P.~M. acknowledges the funding agency DST-INSPIRE, Government of India, for providing the SRF fellowship. R.~P.~S. acknowledges the Science and Engineering Research Board, Government of India, for the Core Research Grant No. CRG/2023/000817.

\nocite{*}
\bibliographystyle{revtex}
\bibliography{Library}
\end{document}